# Multi-imaging and Bayesian estimation for photon counting with EMCCD's

Eric Lantz, Jean Luc Blanchet, Luca Furfaro and Fabrice Devaux,
Université de Franche Comté, Institut FEMTO-ST UMR 6174 CNRS
Laboratoire d'Optique P.M. Duffieux, Route de Gray, 25030 Besançon cedex France
e-mail: elantz@univ-fcomte.fr

**Abstract**

A multi-imaging strategy is proposed and experimentally tested to improve the accuracy of photon counting with an electron multiplying charge-coupled device (EMCCD), by taking into account the random nature of its on-chip gain and the possibility of multiple photo-detection events on one pixel. This strategy is based on Bayesian estimation on each image, with a priori information given by the sum of the images. The method works even for images with large dynamic range, with more improvement in the low light level areas. In these areas, two thirds of the variance added by the EMCCD in a conventional imaging mode are removed, making the physical photon noise predominant in the detected image.

## 1) Introduction

Charge-coupled devices have been for a long time the most efficient detectors in many imaging applications, because of their high quantum efficiency, linear response and very low dark current for cooled cameras. However, conventional low light level CCD's cannot work in the photon counting regime, mainly because of the read-out noise added when the charge of the detected photoelectrons is converted to an output voltage. At slow readout rates (typically some kHz, i.e. several seconds to read a 512×512 pixels sensor), this noise has a standard deviation of 2.5 electrons at best and increases to 10-100 electrons for MHz readout rates often used in applications.

To overcome this problem, cameras with on-chip gain have been proposed for several years. In these electron multiplying charge-coupled devices (EMCCDs), amplification occurs before reading in a multiplication register containing several hundred of cells. Electrons are shifted from one cell to another with a small probability in each cell of being duplicated, resulting in a high mean gain (typically 1000). Because of this high gain, even the signal generated from a single photon emerges from the readout noise floor with high probability. Moreover, a readout rate of 10 MHz is optimal, opening the possibility of taking up to 30 images per second. However, the gain is stochastic, as in an avalanche photodiode, and it is not possible to assign a precise number of photons to each value of the output signal. It can be demonstrated [1] that dividing the output signal by the mean gain results in adding a Poisson detection noise, called excess noise, having the same amplitude as the photon noise. Since both noise sources are independent, the variance in the detected image becomes twice that of the photon noise.

A. G. Basden et al have proposed [2] a photon counting strategy to partially remove this noise for low light level images. Indeed, for an image recorded with a light level of much less than one photon per pixel during the exposure time, pixels with either one or zero photoelectrons can be distinguished by thresholding, while the probability that two photons were incident on the same pixel can be neglected. Ref. 2 proposes the extension of thresholding to more intense images by using multi-thresholding and demonstrates the removal of more than half of the excess noise until a level of 2-3 photons/pixel. However, this paper considers an ideal EMCCD, where the noise comes only from the random gain. In an actual device, the readout noise and clock induced noise (CIC) must be considered. CIC refers to spurious electrons generated by the operation of transferring signal through the device [3]. For high gains and for a sufficiently cooled camera, CIC predominates over dark noise and readout noise. More fundamentally, Basden et al. apply a "photometric correction" that uses a mean level of light. Clearly, such a correction is possible only if some a priori information is available, either that the level of light is constant in some space direction, for example in a fringe [4], or that several images have been recorded at successive instants with the same repartition of light level. In both cases, the information that is actually retrieved is in fact this mean level: thresholding of individual images (or pixels in the first case) is useful to more accurately determine this mean level but not to determine the level of each individual image. Indeed, these images do not carry individual information, because of the assumption of constant light level, and are only more dispersed replica of this mean level.

We propose in this paper an optimal method to determine pixel per pixel the light level by combining information of a set of images. Constant illumination is assumed during the recording of the set. The method is based on Bayesian estimation on each image after thresholding, by using for Bayesian inference the sum of the images. On the other hand, no a priori information is required but the constant illumination assumption and a model of the measurement process. The dynamic range in the image can be large, the only limit being the saturation of the camera.

The paper is organized as follows. In section 2, we assess an optimal light level to apply thresholding. We use in this section a model of sources of noise derived in Appendix 1. In section 3, we present our method. Simulated results are presented in section 4 and compared with experiment in section 5, and a good agreement is obtained. Section 6 is devoted to comparison with previous work, i.e. ref. 2. Finally, we conclude in section 7.

## 2) Thresholding and light level

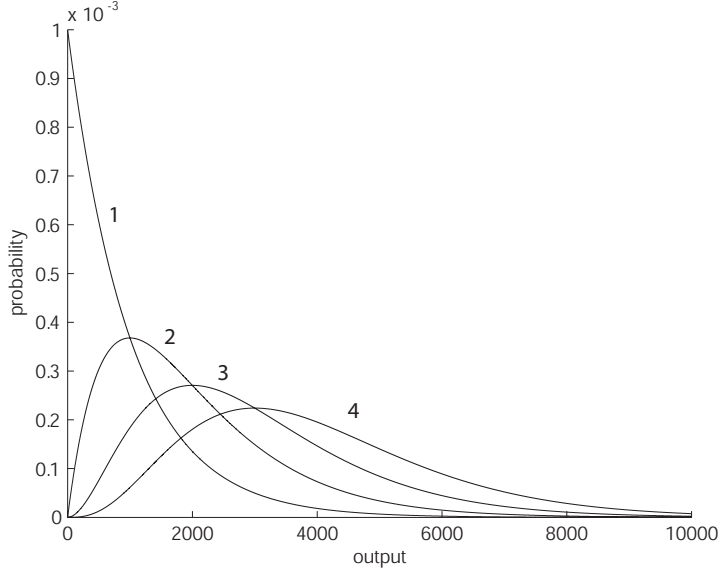

Figure 1: EMCCD output probabilities for a given number of input photons 1,2,3 or 4 and a mean gain of 1000

For n photoelectrons at its input, the multiplication register provides a random output of x electrons, with a conditional probability of x given n [2]:

$$p(x \mid n) = \frac{x^{n-1} \exp\left(-\dfrac{x}{g}\right)}{g^{n}(n-1)!} \qquad (1)$$

where g is the mean gain, equal to $p_c{}^m$. $p_c$ is the probability of duplication of an electron at each cell of the multiplication register and m is the number of cells of this register.

Figure 1 shows these probabilities for 1 to 4 input photoelectrons. It appears clearly from fig. 1 that the inverse problem of finding n given x has no unique solution. In the absence of noise, there is an exception: if the light level is sufficiently low, much less than one photon per pixel in mean, there is either one or zero photon per pixel (from here, we employ "photon" in place of "photoelectron", assuming a unit quantum efficiency of the detector). For a sufficient gain, each photon will give at the output a positive peak, though of random height, while the output will be zero for no input photon. This idealized situation can be extended to higher light levels by acquiring a greater number N of images with an exposure time $\Delta t / N$ instead of an unique image with an exposure time $\Delta t$. N is chosen sufficiently high to ensure much less than one photon/pixel in each individual image.

Things are worse in real world, for two major reasons:

- because of noise, the output is not zero in the absence of photons. Fig A1 in Appendix 1 shows that the probability p(x | 0) is a gaussian centred on zero due to read-out noise, followed by a long tail due to clock induced charges (CIC).

- The above exposed strategy supposes that we know at least an order of magnitude of the incoming light to choose N.

We leave this latter problem for the next section and first show how to optimize N for an approximately known level of light. On each image, assumed to satisfy the assumption of a mean per pixel much less than one, we determine a threshold T and decide there is an input photon if the output x (in grey levels and no in electrons, see Appendix) is greater than T, and else there is no photon. Let y=1 or 0 be the obtained result. Three cases lead to an error:

- There is one input photon and the random gain is sufficiently low so that x≤T. Let $p_{01}$=p(y=0 | n=1) be the corresponding probability.

- Because of noise, x>T for n=0, with probability $p_{10}$.

- There are two photons on one pixel. For a reasonably low threshold, y will be almost always one: $p_{12} \cong 1$.

By choosing N, we can choose the approximate light level $\bar{n}$/N in each image, for a given mean light level $\bar{n}$ on the sum. Since $\mu = \bar{n}$/N is low, the statistic of photons can be considered as Poisson, even if the statistic in the sum image is not. $\bar{n}$/N possesses an optimum: if $\mu$ is too small, the final error will increase because of the noise sources proportional to the number of images; if $\mu$ is too high, the error due to $p_{12}$ increases. More quantitatively, let be Q the quadratic error for one pixel on the sum of the images:

$$Q = N[p_{01}\mu + p_{10}(1-\mu) + p_{12}\mu^2/2] = \bar{n} \ [1-\exp(-T/g)] + (N-\bar{n})p_{10} + \bar{n}^2/2N \qquad (2)$$

$p_{01}$ has been determined by integrating eq.1 until T, for n=1. $p_{10}$ cannot be expressed in compact form (see Appendix 1) but is a decreasing function of T. Both T and N must be optimized to minimize Q. The first term of eq.2 depends only of T, the second term depends of T and N and the third term depends only of N, giving two equations with two unknowns. The solution is, 1-exp(-T/g) being approximated by T/g and N-$\bar{n}$ being approximated by N:

$$T \text{ such that } p_{10} = \frac{g^2(\partial p_{10} / \partial T)^2}{2}, \ N \text{ such that } \frac{\bar{n}}{N} = \sqrt{2p_{10}} \qquad . \qquad (3)$$

By measuring experimentally $p_{10}$ (see Appendix 1), we determine $\frac{\bar{n}}{N}$ = 0.15 photons/pixel and g/T=7, i.e. T=11 grey levels with our camera. Hence an unique image with $\bar{n}$ photons/pixel is at best measured by replacing this image by N images such that the mean level in an image is about 0.15 photons/ pixel. The three errors have then a probability:

$$\mu \ p_{01} = 2.2\%, \ (1-\mu)p_{10} = 0.9\%, \ \mu^2/2 = 1.1\% \qquad (4)$$

It can be noted that T corresponds to 2.8 standard deviations of the read-out noise: with this value of T, $p_{10}$ is due essentially to CIC, while a lower T would result in a rapid increase of errors due to read-out noise.

### 3) Bayesian estimation of photon numbers

A pure thresholding method, as outlined in the previous section, would work for a relatively small range of intensities around the optimal intensity of 0.15 photons/pixel. In other words, we precisely measure the photon number only if we know a priori this photon number, at least approximately, which seems of only limited use. We show in this section how to extend the dynamic range. The proposed strategy is as follows:

- take N (at least 10) images and calculate the sum image.

- Apply the thresholding proposed in the previous section to each individual image.

- Calculate an optimal Bayesian estimator on each pixel of each image, by using as a priori information the value of the corresponding pixel of the sum image.

- Sum these estimators to obtain the best estimation of the sum image.

In order to detail the penultimate step, we first write the Bayes theorem in our particular case. It gives on each pixel of an individual image the probability of n input photons giving a measurement after thresholding y=0 or 1.

$$p(n \mid y) = \frac{p(y \mid n)\, p_{prior}(n)}{p(y)} = \frac{p(y \mid n)\, p_{prior}(n)}{\sum_{k=0}^{k_0} p(y \mid k)\, p_{prior}(k)}, \; n = 0,...,k_0 \qquad (5)$$

Where:

- $k_0$ is chosen to be sufficiently large to allow a great dynamic range in the image. The only limit is that $k_0\, g$ must be below the limit allowed by the analogue to digital conversion.

- $p_{prior}(n)$ obeys a Poisson law with a mean $\mu$ deduced pixel per pixel from the sum image, i.e., if j is an image number:

$$\mu = \frac{\sum_{j=1}^{N} x(j)}{N.g}, \quad p_{prior}(n) = \frac{\exp(-\mu)\mu^n}{n!} \qquad (6)$$

- p(y|n) has been studied in the previous section for n=1 or 0. For n>1, it can be stated that $p(1 \mid n>1) \cong 1$. Slight further precision can be gained as $p(y=1 \mid n \geq 1) = 1 - (\exp(-T/g))^n$ and $p(y=0 \mid n \geq 1) = (\exp(-T/g))^n$

From eq.5, the Bayesian estimator for the $j^{th}$ individual image is simply obtained as :

$$\hat{n}(j) = \sum_{n=1}^{k_0} n.p(n \mid y) \qquad (7)$$

and finally for N images :

$$\hat{n}_{tot} = \sum_{j=1}^{N} \hat{n}(j) \qquad (8)$$

Let us stress that only this final result makes sense: the whole algorithm is based on the hypothesis that the intensity is constant during the exposure time of the N images. As stated in the introduction, each individual image contains less information than this sum, except if the constant intensity hypothesis is broken. In this latter case, there is no clear a priori information available for each image. (See also section 6 for a discussion of this point with respect to the point of view of ref. 2).

### 4) Numerical results

We can expect from section 2 that the best results are obtained with the algorithm of section 3 when the total number of photons per pixel $n_{tot}$ is of the order of 0.15 N. However, the dynamic range is preserved: results are much better than a simple proportional detection for a wide range around this optimal value and there is no superior limit to the validity of the algorithm: for high $n_{tot}$, it tends to the results corresponding to a proportional detection since thresholding of individual images gives always one and the posterior probability becomes the a priori probability, proportional to grey levels. The transition is smooth and some improvement remains for intermediate light levels.

We prove these assertions by studying the quadratic error when measuring a continuous background with a mean summed level $l$ expressed in photons. If the EMCCD mean gain is known, this level can be measured experimentally with a good precision by averaging a great number P of pixels in an image area where the level is as constant as possible. We numerically obtained images by first generating N.P numbers obeying a Poisson distribution of mean $l/N$, then by simulating the amplification and reading of these photons by the EMCCD. We define the efficiency $E$ of our method as:

$$E = \frac{\overline{(\hat{n}_{tot} - l)^2} - l}{l} \qquad (9)$$

where over line means averaging on the P pixels and $\hat{n}_{tot}$ is defined for each pixel by eq. 8 from N individual images of level $l/N$. $E$ characterizes the quadratic error added by the proposed measurement process using N images, compared to the error induced by proportional detection on a single image with an exposure time N times longer. The subtraction of $l$ from the numerator of eq.9 removes the part of the error due to the physical photon Poisson noise: $E$ =0 means a perfect detection while $E$ =1 means no improvement with respect to proportional detection.

Fig. 2 shows, from numerical simulation with N×P=2×10$^5$, the values of $E$ for $l/N$ varying from 10$^{-2}$ to 10 photons and N equal to 5, 10 or 20. With our method, more than half of the quadratic error is removed on a wide range of light levels, from $l$=2×10$^{-1}$ until $l$=5, provided that at least 10 images are acquired. More images are not necessary provided that $l/N$ remains smaller than 0.5. As expected, the method tends to usual results if $l/N$>~1, though some improvement still remains until $l/N$=2. The optimum $l/N$ is 0.1, approximately as expected from section 2. Finally, we have verified that the quadratic error induced by proportional detection is equal to twice the mean level, meaning that the numerator in eq. 9 could be replaced without any noticeable change by $\overline{(\hat{n}_{tot} - l)^2} - \overline{(n_{proport} - l)^2} + l$, if $n_{proport}$ is the result of the proportional method on images of level $l$. This replacement will be performed when dealing with experimental results (next section).

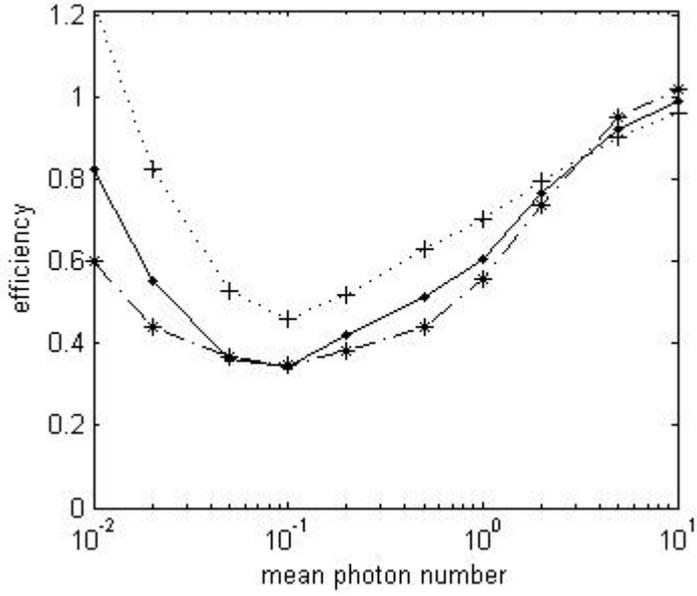

Fig. 2: efficiency (defined in text) versus the mean photon number $l$/N
per pixel and per individual image.
Dotted line: 5 images, Solid line: 10 images, Dash-dotted line: 20 images.
In this figure and the following, lines are an aid for eye and simulated or measured values correspond to the
marked points.

In the preceding paragraphs, we have defined the efficiency by assessing the differences between the simulated measurements and a known mean level, in order to compare the simulation to experiments (see section 5). Numerically, it is also interesting to assess the difference between simulated measurements and the true photon number input values, though no comparison with experiment is possible. We define the misfit function $M$ as [2]:

$$M_{tot} = \frac{\overline{(\hat{n}_{tot} - n_{tot})^2}}{\overline{n}_{tot}}, \qquad M_{indiv} = \frac{\overline{(\hat{n} - n)^2}}{\overline{n}} \tag{10}$$

where the subscripts tot and indiv refer to respectively, differences on the sum of N images or on individual images, with in both cases a priori information taken pixel by pixel from the sum of N images. To calculate $M_{indiv,}$ averaging is performed on the N images as well on the pixels.

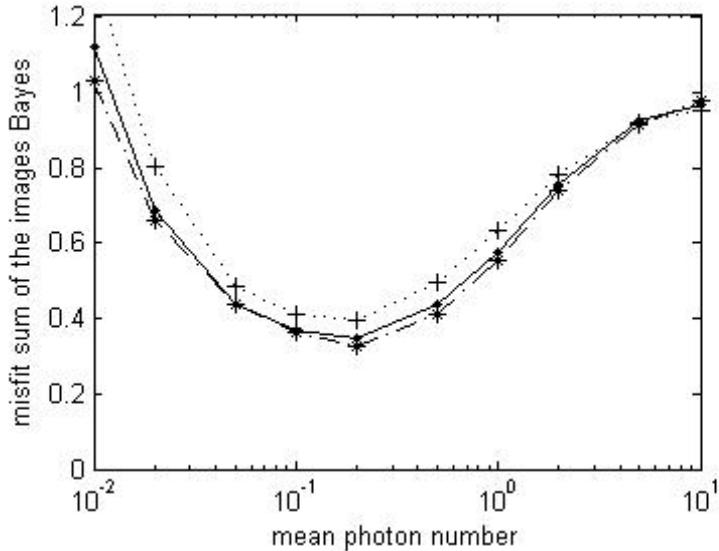

Fig. 3: misfit function (defined in text) on the sum of the images versus the mean photon number per pixel and per individual image (same scales and symbols as in fig. 2).

Fig. 3 shows the values of $M_{tot}$, that are similar to that of $E$ : the quadratic error with respect to a given light level $l$ is indeed the sum of the photon noise $l$ plus the detection noise $M \times l$.

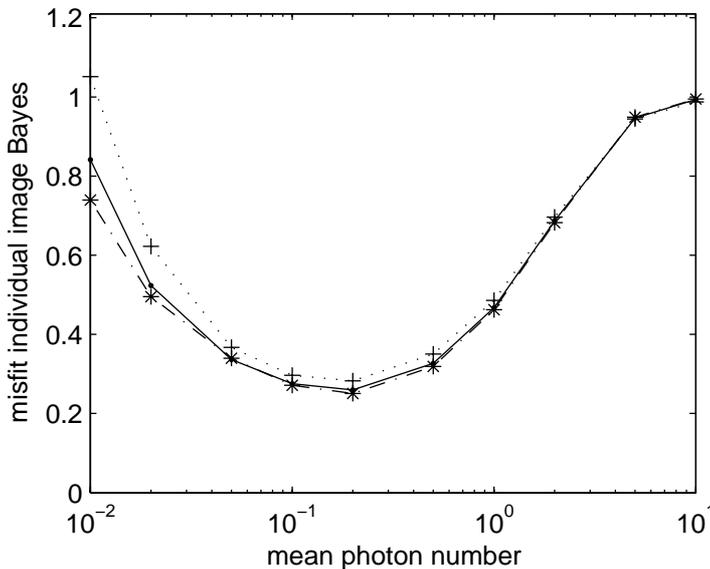

Fig. 4: Misfit function on the individual images (same scales and symbols as in fig. 2)

As mentioned above, there is no clear supplementary information in individual images, because a fundamental assumption is that all images have the same light level on a given pixel. Nevertheless, we present the misfit function for individual images in fig. 4. Though similar, the results are somewhat better than on the sum image.

Results for individual images can be improved: thresholding can be replaced by the complete determination of the posterior probability p(n|x), where x is the number of grey levels of one pixel of one image. Eq. 5 to 7 remain valid, if the binary value y= 0 or 1 is replaced by x. Before applying these equations, p(x|n), n=0,....,$k_0$ must be calculated by convolving the probability p(x|0) determined in Appendix 1, representing the noise, by the

probability p(x|n) in the absence of noise, given by eq. 1. Fig. 5 shows the obtained misfit function: its value is divided by 2 for high levels. Note however that the same technique (called full posterior method in the following) applied to the misfit on the sum image (first part of eq. 10) shows no improvement with respect to the Bayes thresholding technique. This latter method is more rapid ($4 \times 10^{-6}$ seconds per pixel and per image with the thresholding method versus $30 \times 10^{-6}$ seconds with the full posterior method, in Matlab on our computer Dell with an Intel processor Xeon at 2.8 GHz).

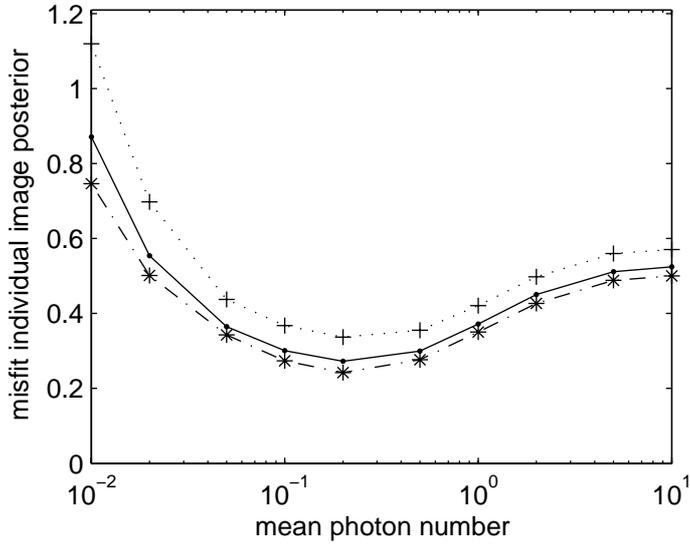

Fig. 5: Misfit function on the individual images: method of posterior probability without thresholding (same scales and symbols as in fig. 2)

**5) Experimental results**

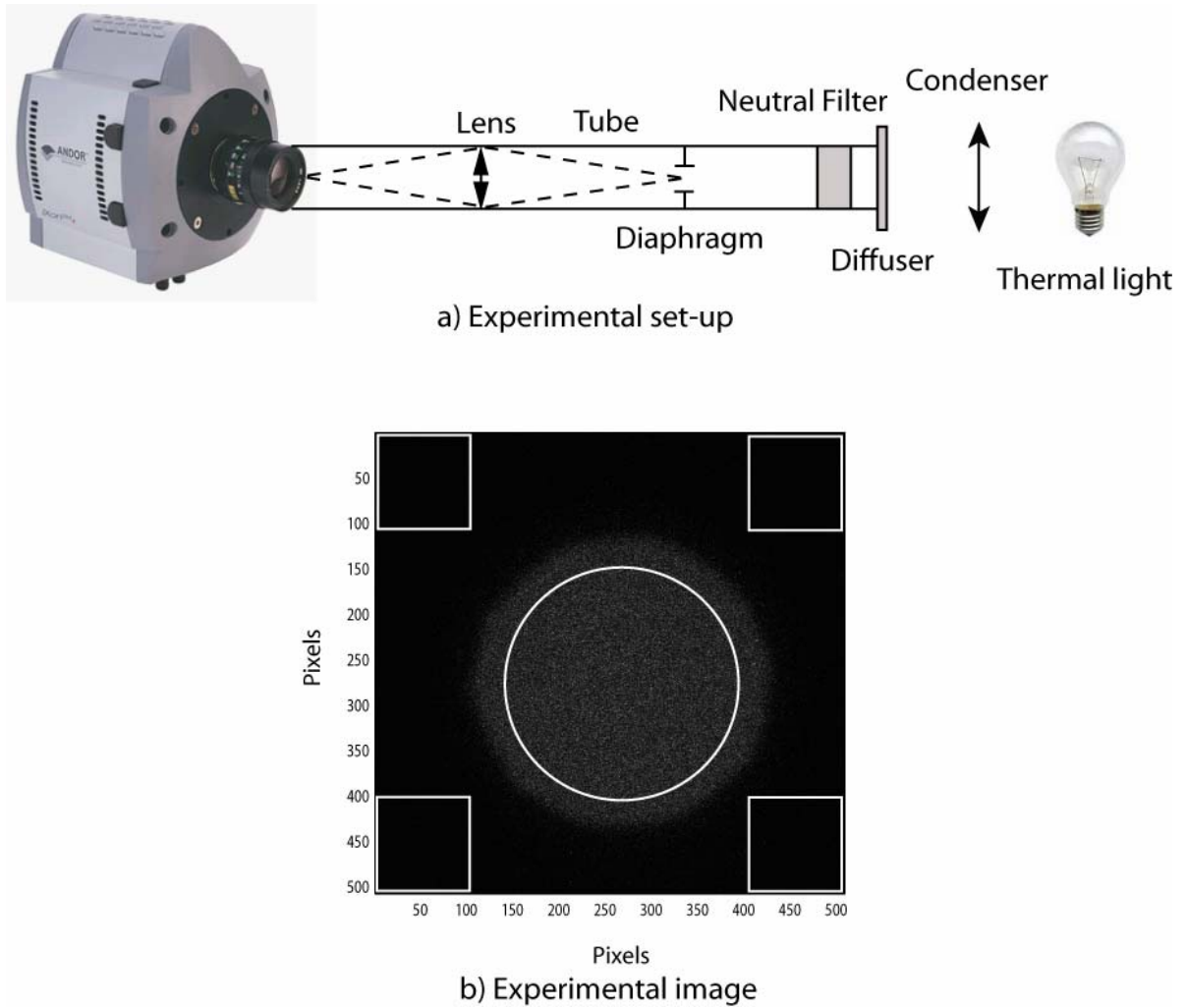

a) Experimental set-up

b) Experimental image

Fig. 6: (a) Experimental set-up. (b) Experimental image: the statistics are performed inside the circle and the mean square plane is determined with the pixel values inside the squares.

Fig.6 shows the experimental set up. The incoming light, provided by a halogen lamp, enlightens a diaphragm which is imaged onto the camera. An optical neutral density filter, a condenser and a diffuser ensure a low-level highly homogeneous illumination of the diaphragm. All the trajectory of the light after the diaphragm is enclosed in a tube in order to avoid parasitic reflections.

The camera is a back-illuminated Electron Multiplying CCD from Andor technology, model iXon+ DU897-ECS-BV [5]. The quantum efficiency is greater than 90 % for wavelengths in the visible range. The detector area has $512 \times 512$ pixels, with a pixel size of $16\mu m \times 16\mu m$. We used a readout rate of 10 MHz at 14 bit and the camera was cooled at -85°C. We acquired N images with an exposure time $\Delta t/N$=33ms and compared the result with proportional detection on one image with an exposure time $\Delta t$. For example $\Delta t$=330ms for N=10. The EM nominal gain was set to 1000.

To estimate the mean intensity, the electronic background must be carefully subtracted. Results obtained in darkness with the shutter closed (see Appendix 1) show that this background has not exactly the same value all over the detector area. This very weak difference is of the order of one grey level, i.e. 1/100 photon after division by the gain. It must nevertheless be compensated to obtain results for images with light levels of few hundredths of photons. Therefore a least mean squares plane has been calculated on each image by using only the grey level values from non-illuminated square areas on the four corners of the image: see figure 6b. This plane has been removed from the images to obtain a quasi perfect zero mean in the non-illuminated areas.

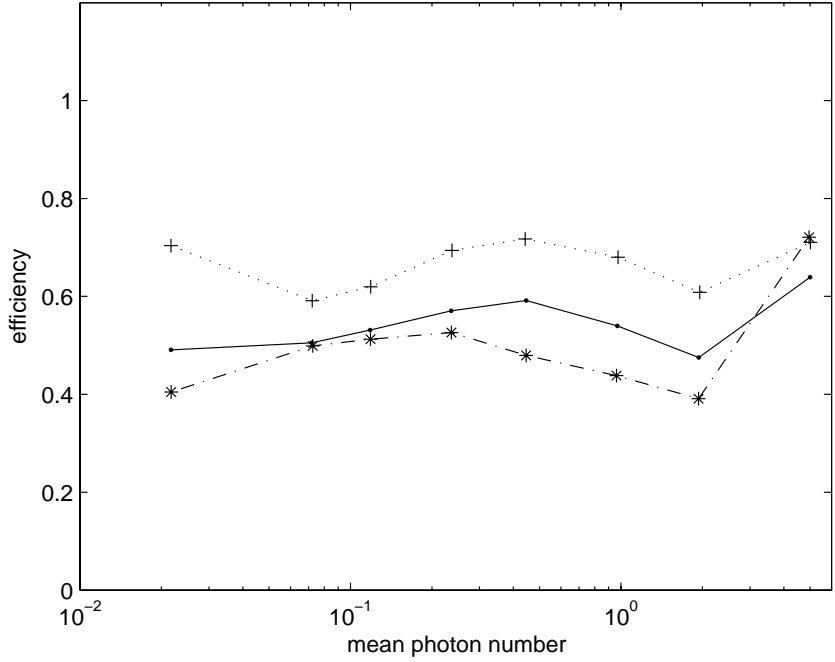

Fig. 7: Experimental efficiency versus the mean photon number per pixel and per individual image. Dotted line: 5 images, solid line: 10 images, dash-dotted line: 20 images.

To determine the experimental efficiency, a circular area of $3.2 \times 10^4$ pixels has been selected, where there is no evident deterministic variation of the intensity. In this area, we have, for a flux of 0.2 photons/pixel and a nominal gain of 1000:

$$\overline{(n_{proport} - l)^2} = 2l + \sigma_{nolight}^2 \qquad (11)$$

Where $\sigma_{nolight}^2$ is the variance in absence of incoming light, determined in Appendix 1. For higher light levels, the variance is higher than twice the mean and a supplementary term $\sigma_{\sup pl}^2$ must be added (its value is for example 0.36 for $l$=9.8 photons per pixel; see below some comments about its origin) and, to compare our method with proportional detection on one image, we assume:

$$\overline{(n_{proport} - l)^2} = 2l + \sigma_{nolight}^2 + \sigma_{\sup pl}^2 , \qquad \overline{(\hat{n}_{tot} - l)^2} = l + \sigma_{nolight}^2 + \sigma_{\sup pl}^2 + \sigma_{Bayes}^2 \qquad (12)$$

Where $\sigma_{Bayes}^2$ is the quadratic error added by the random gain of the EMCCD after estimation with our method. Hence, the efficiency of our method can be experimentally determined as:

$$E = \frac{\sigma_{Bayes}^2}{l} = \frac{\overline{(\hat{n}_{tot} - l)^2} - \overline{(n_{proport} - l)^2} + l}{l} \qquad (13)$$

Fig. 7 shows the experimental values of $E$ for $l$/N varying from 0.02 to 5 photons and N equal to 5, 10 or 20, in rather good agreement with simulation (fig.2). Actually, experimental results are better than the simulation for $l$/N=1 or 2, because the performance of our EMCCD appears, whatever the detection algorithm, to be better at this level than for a N times higher level: see below. Note that $l$ is determined by averaging the proportional values and not the Bayesian estimator, because this Bayesian estimator is biased. See Appendix 2 for some comments on advantages and precautions of use of biased estimators.

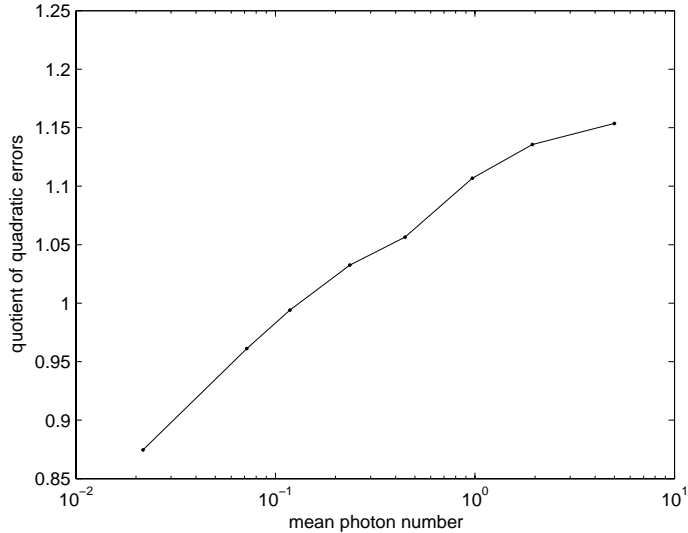

Fig. 8: Pure proportional detection: quadratic error for a single image of level $l$ divided by the quadratic error for 10 images of level $l/10$, versus the level $l/10$.

Surprisingly, a level of 0.2 photons per pixel per individual image seems optimum even for pure proportional detection. Fig. 8 presents a comparison of quadratic errors with proportional detection resulting either from the detection of one image of level $l$ or from the sum by software of 10 images of level $l/10$. The quadratic error is smaller for the multi-imaging strategy as soon as the level $l/10$ on each individual image is greater than 0.2 photon per pixel. It means that some supplementary noise, called above $\sigma^2_{\sup pl}$, is added when the signal level increases. This supplementary noise appears of higher level than the noise due to readout even at relatively low intensities.

## 6) Comparison with previous work

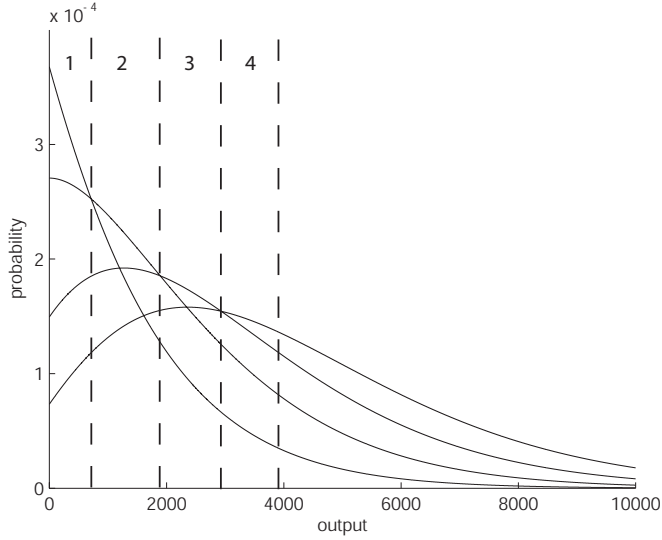

Figure 9: Model without noise: output probability distribution for a given integer light level $l$/N and threshold boundaries for the multi-thresholding strategy proposed by Basden et al.

It can be seen in fig. 9 that for each value x of the output of a perfect EMCCD, an integer value of the light level $l$/N gives the highest probability p(x|($l$/N)): corresponding domains are labelled by $l$/N on Fig. 9. Basden et al. have proposed [2] a multi-thresholding strategy where this value is retained as the estimator $I_{est}$ of n given x. As is, this estimator is biased: for a light level $l$, mean of a Poisson distribution of the input photons, the mean of the estimator is greater than $l$ for $l$<20 photons/pixel. Ref. 2 proposed to correct this estimator according to:

$$I_{corrected} = \frac{I_{est}}{1 + 0.7 \exp(-I_{est}/3)} \tag{14}$$

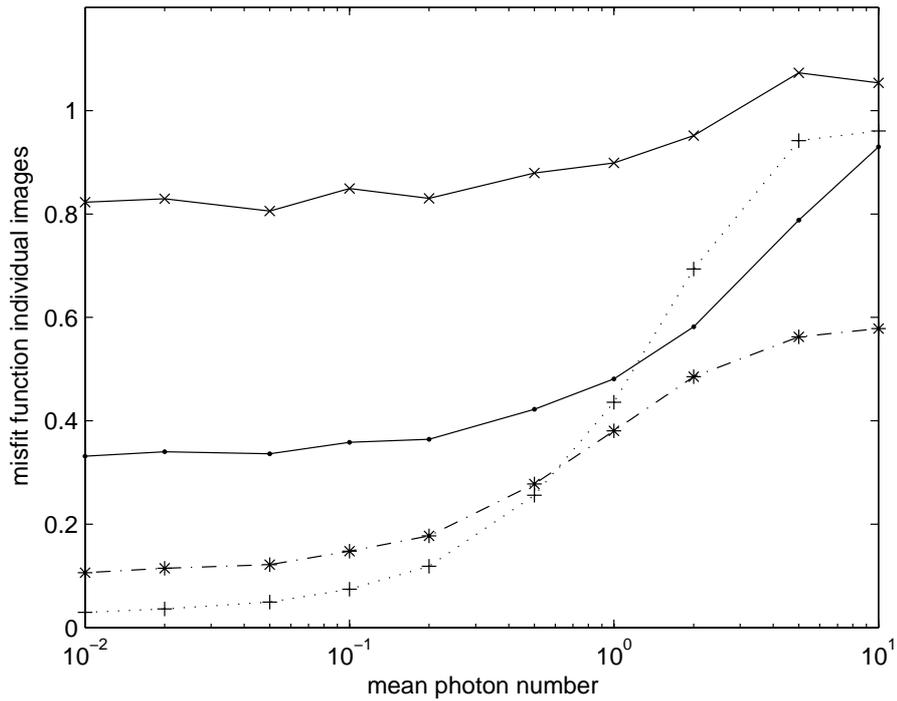

Fig.10: misfit function on the individual images versus the mean photon number per pixel and per individual image, for 10 images, model without noise.
Dotted line: Bayes thresholding, Dash-dotted line: full posterior method. Solid line with points: Basden et al method with level of photometric correction determined on the 10 images. Solid line with x: photometric correction determined only from one image.

It can be easily verified from eq.1 that this corrected estimator is nonbiased, with a high precision, if $I_{est}$ is the mean of a great number of images. However we argue (see introduction) that this situation is not very interesting, because this mean carries more information than the individual images. Nevertheless, we show in fig. 10, for comparison, the misfit function of the individual images when both our a priori information and the photometric correction of ref. [2] are calculated from 10 images (results are almost identical for more images). To allow a direct comparison with [2], the same model of EMCCD, given by eq. 1, has been used, with negligible readout and CIC noise. Results are clearly better with our method, at least for fluxes smaller than one photon per pixel. For higher fluxes the full posterior method seems preferable.

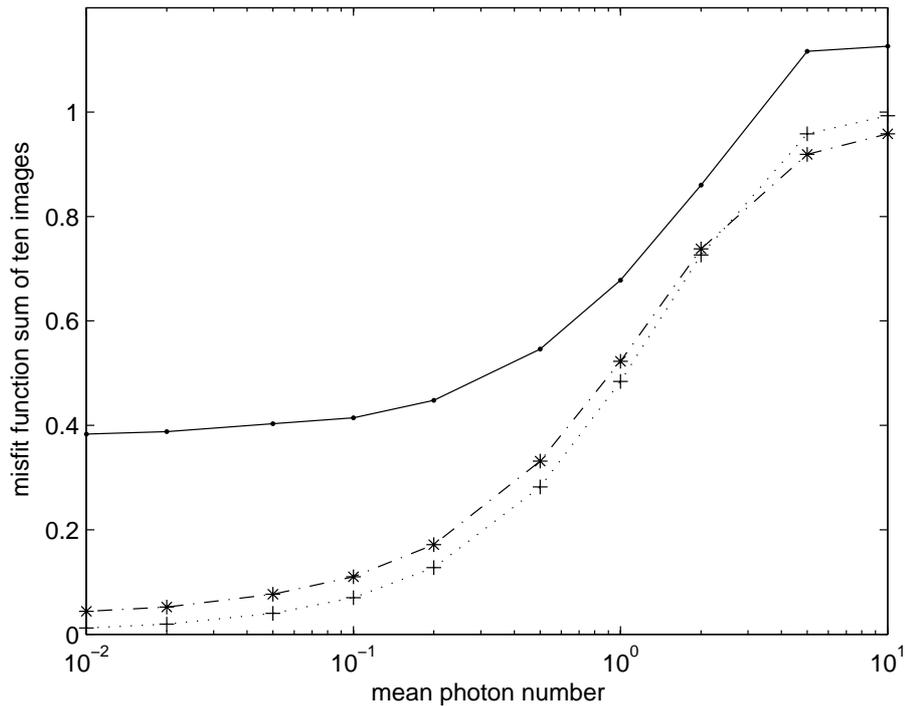

Fig. 11: misfit function on the sum image versus the mean photon number per pixel and per individual image, for 10 images, model without noise.
Dotted line: Bayes thresholding, Dash-dotted line: full posterior method. Solid line with points: Basden et al method with level of photometric correction determined on the 10 images.

This conclusion does not hold if the misfit function is calculated on the sum of the 10 images. Fig. 11 shows the misfit function for this sum image: the Bayes thresholding method appears to be the best method for low light levels, and similar to the full posterior method at higher levels. It could appear surprising that the full posterior method does not give the best results in all cases: indeed, it is well known that the mean of the posterior probability is the estimator that minimizes the quadratic error. However in our case the "a priori" probability comes also from the measurements, that rends this theoretical result not applicable.

To conclude this section, the Bayes thresholding method is much more efficient than the other methods at low light levels and converges smoothly to the results of proportional detection for levels as high as 10 photons per pixel per individual image, or 100 photons per pixel on the sum. Whatever the level, this method is either more efficient or as efficient as the others.

## 7) Conclusion

EMCCD's allow efficient imaging at very low light levels. With proportional detection, the random gain results in doubling the photon Poisson noise, while conventional cooled slow-scan CCDs add a readout noise of 2.5 photoelectrons rms. Hence, with proportional detection, EMCCD's are more efficient than conventional CCDs only for light levels smaller than approximately 6 photoelectrons per pixel. With a commercially available EMCCD, our strategy allows this limit to be pushed to 12 photoelectrons per pixel by at least halving the added noise. The limit is far higher in applications where rapid acquisition is needed, like "lucky imaging" [6]. For very low light levels, the performances are better described in terms of equivalent quantum efficiency: proportional detection results in dividing by 2 the equivalent quantum efficiency while this quantum efficiency is divided by 1.5 or less with our algorithm. Improvements will be more spectacular when EMCCD's will become closer of their ideal performances, i.e. random gain but negligible other noise sources: with such an ideal EMCCD, we have proved in section 6 that the noise added with our algorithm tends to zero at low light levels, while this noise tends to the usual Poisson noise at high light levels, allowing a high signal to noise ratio to be obtained in the entire image, whatever its dynamic range (noise refers here to the added noise and not to the unavoidable photon noise). It should be noted that some groups are developing for astronomical purposes EMCCD with higher gains than ours, leading to performances of our algorithm intermediate between the ideal case and our experimental results [7]. Moreover, in the present state of technology, multi-imaging seems useful even without using our algorithm, because the variance of the added noise increases more rapidly than the light level. Actually, new methods using EMCCD in astronomy, like lucky imaging [6], are multi-imaging in essence. In lucky imaging, the a priori hypothesis of constant or quasi-constant illumination between images is not fulfilled in the original images, because of random distortions due to atmosphere. Hence, it could be preferable to first use proportional images to determine the shifts to apply [8], then to use our algorithm on the set of selected shifted images to improve the photometry. This very promising perspective deserves future investigation.

## Appendix A

We present in this Appendix a quantitative model of sources of noise affecting an EMCCD camera. The most important ones are the read-out noise, the clock-induced noise (CIC), i.e spurious electrons generated during the transfers even in absence of input photoelectrons, and the dark-noise, i.e. charges generated in the pixels in absence of light. The probabilities associated at each noise source can be described as follows:

- readout noise : it is well  described by a Gaussian probability with a zero mean and a standard deviation $\sigma_{read.}$

- cic and dark noise. There is a small probability $p_{par}$ that a spurious electron is present at the input of the multiplication register when reading a given pixel. Most of these electrons are generated during the parallel transfer [3] , the dark noise being negligible for short exposure times. Then, the probability of $x_{par}$ electrons be present at the output is given by, using eq. 1:

$$p(x_{par})=p_{par}\cdot p(x_{par}|1)=\frac{p_{par}\ \exp\left(-\dfrac{x_{par}}{g}\right)}{g} \qquad (A1)$$

- There is a small probability $p_{ser}$ that a spurious electron is generated at each cell of the serial multiplication register, giving at the output:

$$p(x_{ser})=\sum_{l=1}^{m}\frac{p_{ser}\exp\left(-\dfrac{x_{ser}}{p_c^{m-l}}\right)}{p_c^{m-l}} \qquad (A2)$$

Formulas (A1) and (A2) are not valid for x=0, that corresponds (fortunately!) to the biggest probability. $p(x_{par}=0)$ is simply obtained as $1- p(x_{par}>0)$, after calculation of p(x) until sufficiently high values of x. The procedure is identical for $p(x_{ser})$.

p(x|0) is obtained by convolving the three probability curves. Note that negative values of x are possible after deduction of the electronic offset of the camera, because of the read-out noise centred in zero. The last step consists in converting x electrons in $x_g$ grey levels (gl). For our camera, a grey level corresponds to 11.9 output electrons.

We have estimated the level of these different noise sources for our EMCCD camera Andor, model IXION, by taking a set of images in darkness with the shutter closed and an exposure time of 33ms. The sensor was cooled at -85°C and the camera was operated with a pixel read-out rate of 10 MHZ and a vertical c<lock speed of one shift/0.5μs. The theoretical probability law described above was fitted to the histogram of pixels using a Gauss-Newton algorithm: fig. A1. The fitting gives $p_{par}$=3.9±0.3×10$^{-3}$, $p_{ser}$=4.1±0.6×10$^{-5}$, $\sigma_{read}$=46.2±0.1 electrons. The experimental variance of 94 gl$^2$ is the sum of 15 gl$^2$ due to the read-out noise, 21 gl$^2$ due to the serial noise and 58 gl$^2$ coming from the parallel transfer ($\sigma^2_{par} \cong 2\ g^2\ p_{par}$). These results are in agreement with the technical notice of e2v which states that the CIC from the parallel transfer dominates [3] and with the results of Daigle al [9], who give a mean CIC of the order of 2.6×10$^{-3}$ e$^-$/pixel/frame. Note also that the total standard deviation of 9.7gl can be expressed as 0.12 photons, after conversion of the grey levels in output electrons and division by the gain.

.

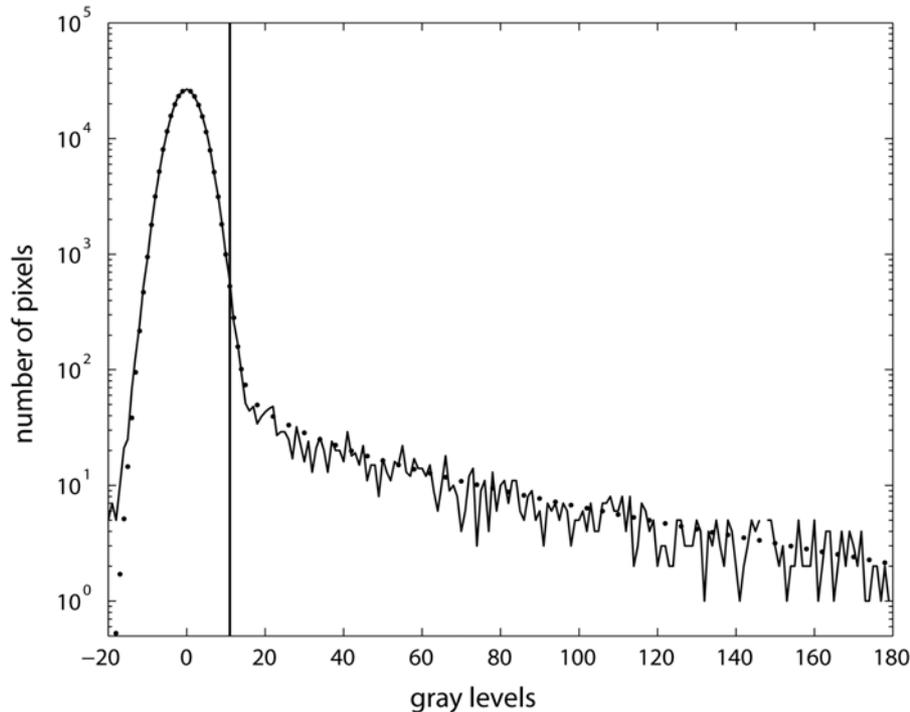

Fig A1: Histogram of grey levels for a dark image. Full line: experimental values. Dotted line: fitting with the model described in the text. The vertical line corresponds to the decision threshold at 11 grey levels (see sections 2 and 3).

Fig A1 presents the experimental results fitted by the model. The threshold at 11 grey levels introduced in section 2 corresponds to 11×11.9/46.2=2.8 standard deviations of the read-out noise.

**Appendix B**

The estimator used in this paper is biased: $\langle \hat{n}_{tot} \rangle \neq l$, where $\langle \ \rangle$ designates the mean in its mathematical sense of expected value. It means that if, using each time N images, we can repeat the measure on an infinite number P of pixels illuminated by the same light level, the mean of the proportional estimator would give the true value $l$, while our estimator does not. Note however it is convergent: $\lim_{N \to \infty, \, l/N \, constant} (\hat{n}_{tot} / N) = l / N$. We could easily prove these assertions. We prefer to give an example that will help, we hope, to understand the origin of this situation.

The simplest situation arises when we can assume with a good confidence that there is either one or zero photon per pixel on each individual image, for example for $\mu = 10^{-2}$, where $\mu$ is defined in eq.6. After thresholding, we want to replace the obtained outputs y = 0 or 1 by values $\hat{n}(0)$ and $\hat{n}(1)$ calculated in order to either obtain a global non biased estimator $\hat{n}$, for N=1, or to minimize the quadratic error between the true value and the estimator. The two objectives lead to very different values $\hat{n}(0)$ and $\hat{n}(1)$.

To obtain a non biased estimator, we must have, $p_{ij}$ being defined as $p(y = i \mid n = j)$:

$$\langle (\hat{n} \mid n = 0) \rangle = 0 = p_{00}\hat{n}(0) + p_{10}\hat{n}(1)$$
$$\langle (\hat{n} \mid n = 1) \rangle = 1 = p_{11}\hat{n}(1) + p_{01}\hat{n}(0) \tag{B1}$$

The solution of eqs. B1 gives $\hat{n}(0) < 0$ and $\hat{n}(1) > 1$. Intuitively, it seems strange to take into account the possibility of having a true n=1 when detecting 0 by replacing 0 with a negative value. Indeed, such a strategy increases the mean error (in absolute value, or quadratic) between the estimator and the true value. It is justified only if we make many measurements of the same true value. It is not the case if we want a wide dynamic range to be allowed in the image.

To use the Bayesian estimator of eq. 7, we must give values to $p_{prior}(n)$ used in eq. 5. Once done, eq. 5 and 7 give, with the assumptions of this Appendix:

$$\hat{n}(1) = \frac{p_{11}p_{prior}(1)}{p_{11}p_{prior}(1) + p_{10}p_{prior}(0)}, \quad \hat{n}(0) = \frac{p_{01}p_{prior}(1)}{p_{01}p_{prior}(1) + p_{00}p_{prior}(0)} \tag{B2}$$

Both values lie between 0 and 1, whatever the a priori probability. Note also that $\hat{n} = 1$ whatever the measurement if $p_{prior}(1) = 1$: if the a priori information is stronger than the measurement, it gives the final result. We encounter a similar situation for the strategy developed in section 3 when there is much more than one photon per pixel in the individual images: thresholding gives almost always one and all the information comes from the proportional measurement.

To summarize this Appendix, a non biased estimator could give values far away from the true value with a great probability, even if the mean of this great error is zero. In practice, a non biased estimator should also be efficient: its variance is minimum, among all non-biased estimators. Eqs (B1) define of course an efficient estimator, because it is the only one that is non-biased. We propose to use a biased estimator, closer for each measurement of the true value although its mean is not equal to this true value. Despite its advantages, this estimator must be employed with caution: for example, the best estimator of an average on P pixels is not the average of the estimators on one pixel, but rather the estimator formed by using the N×P available pixel values altogether.

**Acknowledgement**

This work has been supported in part by the Agence Nationale pour la Recherche, project IRCOQ.

**References**

[1] J. Hynecek, T. Nishiwaki, "Excess noise and other important characteristics of low light level imaging using charge multiplying CCDs", IEEE Trans. of Electron devices **50,** 239-245 (2003)

[2] A. G. Basden, C.A. Haniff, C. D Mackay, "Photon counting strategies with low-light level CCDs", Mon. Not. R. Astron. Soc. **345,** 985-991 (2003)

[3] Low-Light Technical Note 4, "Dark Signal and Clock-Induced Charge in L3Vision CCD Sensors", e2v technologies, at http://www.e2v.com/products/ccd-and-cmos-imaging-and-semiconductors/imaging-l3vision/

[4] A. G. Basden and C. A. Haniff, "Low light level CCD's and visibility parameter estimation", Mon. Not. R. Astron. Soc **347** 1187-1197 ( 2004)

[5] Andor Catalog, at www.andor.com

[6] N.M. Law, C.D. Mackay, and J.E. Baldwin, "Lucky Imaging: High Angular Resolution Imaging in the Visible from the Ground" A&A **446**, 739-745 (2006)

[7] A. Basden, private communication. To the best of our knowledge, no assessment of performances of such a camera has been yet publicly available.

[8] A.S. Fruchter and R.N. Hook, "Drizzle: A Method for the Linear Reconstruction of Undersampled Images", PASP **114,** 144-152 (2002)

[9] O. Daigle, J.L. Gach, C. Guillaume, C. Carignan, P. Balard , O. Boissin, "L3CCD results in pure photon counting mode", Proc. of SPIE **5499**, Optical and Infrared Detectors for Astronomy, James D. Garnett, James W. Beletic, Ed., 219-227 (2004)